\documentclass[review]{elsarticle}

\usepackage{amsmath}
\usepackage{amsfonts}
\newcommand{\bs}[1]{\boldsymbol{#1}}

\usepackage{lineno,hyperref}
\modulolinenumbers[5]

\journal{Nowhere}

\begin{document}

\begin{frontmatter}

\title{Fast dynamic aperture optimization with reversal integration}

\author[bnl]{Yongjun Li\corref{correspondingauthor}}
\cortext[correspondingauthor]{Corresponding author}
\ead{yli@bnl.gov}

\author[msu]{Yue Hao}
\author[lbnl]{Kilean Hwang}
\author[bnl]{Robert Rainer}
\author[bnl]{An He}
\author[euclid]{Ao Liu}

\address[bnl]{Brookhaven National Laboratory, Upton, New York 11973, USA}
\address[msu]{FRIB/NSCL, Michigan State University, East Lansing, Michigan
  48864, USA}
\address[lbnl]{Lawrence Berkeley National Laboratory, Berkeley 94720,
  California, USA}
\address[euclid]{Euclid Techlabs, LLC, Bolingbrook, Illinois 60440, USA}

\begin{abstract}
  A fast method for dynamic aperture (DA) optimization of storage
  rings has been developed through the use of reversal
  integration. Even if dynamical systems have an exact reversal
  symmetry, a numerical forward integration differs from its
  reversal. For chaotic trajectories, cumulative round-off errors are
  scaled, which results in an exponential growth in the
  difference. The exponential effect, intrinsically associated with
  the Lyapunov exponent, is a generic indicator of chaos because it
  represents the sensitivity of chaotic motion to an initial
  condition. A chaos indicator of the charged particle motion is then
  obtained by comparing the forward integrations of particle
  trajectories with corresponding reversals, a.k.a. ``backward
  integrations.''  The indicator was confirmed to be observable
  through short-term particle tracking simulations. Therefore,
  adopting it as an objective function could speed up
  optimization. The DA of the National Synchrotron Light Source II
  storage ring, and another test diffraction-limited light source
  ring, were optimized using this method for the purpose of
  demonstration.
\end{abstract}


\end{frontmatter}


\section{\label{sect:intro}Introduction}

  Accurate computation of the Lyapunov exponent (LE) of particle motion in
  accelerators and comparison with numerical dynamic aperture (DA)
  simulations have been well studied. Past examples include
  ~\cite{Zimmermann:1994pz, Habib:1995, Scandale:1997jr,
    Giovannozzi:1997jn, Giovannozzi:1997uc, Turchetti:2018,
    Schmidt:1988,Fischer:1995,Schmidt:1991}.  A general correlation
  between the LE and the DA has been confirmed, but a universal or
  quantitative equivalence has yet to be established. In some studies, the
  LE was found to underestimate the DA in storage
  rings~\cite{Giovannozzi:1997uc}.  Additionally, accurate calculation of
  the LE ~\cite{Wolf:1985, Habib:1995} is time-consuming due to the
  long-term numerical integrations required, making its use difficult in
  direct dynamic aperture optimization.

  The discovery of another indicator of chaos, obtained by comparing
  forward integrations and corresponding reversals (i.e. backward
  integration), can be traced back to the
  1950's~\cite{Cole:1994vc}. The method is also known as ``the
  trajectory reversing method'', and has been widely used to estimate
  stable regions of dynamical systems since then~\cite{Miller:1964,
    genesio1984new, chiang1988stability, loccufier2000new,
    lee2000analysis, jaulin2001nonlinear}.  One of the more recent
  uses of this indicator have been to understand the DA of the
  Integrable Optics Test Accelerator (IOTA) in the presence of space
  charges~\cite{Hwang:2019bdh}.  The indicator is intrinsically
  associated with the LE, because it also represents the sensitivity
  of chaotic motion to an initial condition.  We found that
  implementing just a few turns of forward-reversal (F-R) integrations
  reveal an observable difference using high precision (e.g. 64-bit)
  floats for modern storage rings.  Therefore, the chaos indicator can
  be computed at a faster rate. By combining population-based
  optimization, such as multi-objective genetic algorithm
  (MOGA)~\cite{Deb, Yang:2009, Yang:2011, Li:2016, Li:2018, Wan:2019,
    Liu:2015, Liu:2017} with the trajectory reversing method, a fast
  approach for DA optimization has been developed and demonstrated
  with two examples in this paper.  Tracking-based optimization has
  traditionally been limited by time-consuming tracking simulations.
  The new approach provides a potential solution to using short-term
  tracking simulations to optimize the DA for large scale storage
  rings.

  To further explain this approach, the remaining sections are
  outlined as follows: Sect.~\ref{sect:fb} briefly explains the F-R
  integration as an indicator of chaos.  A H\'{e}non map's chaos is
  studied with this method for proof-of-principle in
  Sect.~\ref{sect:henon}.  In Sect.~\ref{sect:application}, we take
  the National Synchrotron Light Source II (NSLS-II) storage ring and
  another test diffraction-limited light source ring as two examples
  to demonstrate the application of this approach.  A brief summary is
  given in Sect.~\ref{sect:summary}.

\section{\label{sect:fb}Forward-reversal (F-R) integrations}

  In dynamical systems, the Lyapunov exponent (LE) is used to
  characterize the rate of separation of two infinitesimally close
  trajectories. In phase space, two trajectories with initial
  separation $\Delta \bs{z}(0)$ diverge at a rate given by,
  \begin{equation}\label{eq:Lyapunov}
    |\Delta\bs{z}(t)| \approx e^{\lambda t}|\Delta\bs{z}(0)|,
  \end{equation}
  where, $\bs{z}(t)=(x,p_x;y,p_y;s,p_s)^T$ is a vector composed of
  canonical coordinates in phase space at time $t$, and $\lambda$ is
  the LE. The superscript ($^T$) represents the transpose of a
  vector. Bold symbols, such as ``$\bs{z}$'', are used to denote
  vectors throughout this paper. The above rate calculation assumes
  the divergence is treated as a linearized approximation. The rate of
  separation can be different for different orientations of the
  initial separation vector, which yields multiple LEs for a given
  dynamical system. The largest LE of a system is referred to as the
  maximal Lyapunov exponent (MLE), which is defined as,
  \begin{equation}\label{eq:mle_t}
    \lambda=\lim_{t\to\infty }\lim _{\Delta\bs{z}(0)\to \bs{0}}
    \frac{1}{t}\ln\frac{|\Delta\bs{z}(t)|}{|\Delta\bs{z}(0)|}.
  \end{equation}
  Here $\Delta\bs{z}(0)\to\bs{0}$ ensures the validity of the linear
  approximation at any given time. The MLE provides valuable
  information about the dynamical system's predictability.

  In accelerators, it is more practical to use the path length of a
  reference particle $s$ rather than time $t$ as the free variable.
  The trajectory of an arbitrary particle can therefore be described
  as a deviation from a reference particle. For example, the momentum
  offset is denoted as $\delta=\frac{\Delta p}{p_0}$.  After some
  canonical transformations~\cite{Ripken:1985qn}, the time
  $t$-integration can be converted to a path length $s$-integration. A
  new MLE $\lambda_s$ can then be re-defined as,
  \begin{equation}\label{eq:mle_s}
    \lambda_s=\lim_{s\to\infty }\lim _{\Delta\bs{z}(0)\to\bs{0}}
    \frac{1}{s}\ln\frac{|\Delta\bs{z}(s)|}{|\Delta\bs{z}(0)|},
  \end{equation}
  where $\bs{z}(s)=(x,p_x;y,p_y;s-ct,\delta)^T$ are new canonical
  coordinates in the phase space at position $s$, and $s-ct$ is the
  longitudinal coordinate offset. For convenience, the rest of this
  manuscript will use path length $s$ of particle motion as the free
  variable unless stated otherwise.

\begin{figure}[!ht]
    \centering \includegraphics[width=1.\columnwidth]{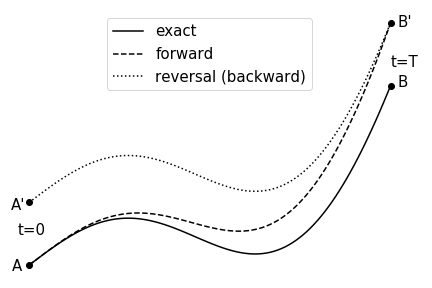}
    \caption{\label{fig:fb} Schematic illustration of forward and
      time-reversal integrations for a dynamical system. The solid
      line represents the exact trajectory from $A$ at $t=0$ to $B$ at
      $t=T$. The dashed line is the numerical integration, which
      becomes $B^{\prime}$ at $t=T$. The difference between $B$ and
      $B^{\prime}$ indicates the chaos, but in practice, $B$ is
      usually unknown. The dotted line is the time-reversal
      integration starting from $B^{\prime}$ and ending at
      $A^{\prime}$. The difference between two initial conditions $A$
      and $A^{\prime}$ is an indicator of chaos of the system for this
      specific initial condition.}
  \end{figure}

  Generally speaking, the calculation of MLEs as defined above in
  Eqs.~\ref{eq:mle_t}-\ref{eq:mle_s}, often cannot be carried out
  analytically. In these cases, the calculation would therefore require
  the use of numerical techniques~\cite{Habib:1995, Wolf:1985}. An
  alternative, empirical method to measure the chaos of a dynamical system
  is to use a reversal integration as suggested in Ref.~\cite{Cole:1994vc,
    Miller:1964, Hwang:2019bdh}. During proof of concept, the properties
  of the system under time symmetry were calculated by letting the system
  evolve through some number of integration steps, then switching the sign
  of the time step and letting the system run backward until the total
  time variable reached zero. On the return to time zero, the changes in
  corresponding velocities and positions were calculated and collated, as
  was the value of the time variable during the change of sign. A new set
  of initial conditions could then be re-established. Due to the
  unavoidable numerical round-off error~\cite{ieee754-2019, Laslett:1957}
  for a chaotic trajectory, the re-established initial conditions deviated
  from the original ones as illustrated in Fig.~\ref{fig:fb}. The
  difference, a.k.a. the consistency error is an indicator of chaos which
  is associated with its LE.

  The principle of the F-R integration approach can be briefly
  outlined as follows: A nonlinear transfer function, denoted as $f$,
  propagates through an $N$-dimensional phase space coordinate
  $\bs{z}=\left(x_1,x_2,\cdots,x_N;p_1,p_2,\cdots,p_D\right)^{T}$
  iteratively.  In a finite-precision computation process, the
  iteration from the $(n-1)^{\text{th}}$ state $\bs{z}_{n-1}$ to the
  next state $\bs{z}_{n}$ reads as:
  \begin{equation}
    \bs{z}_n=f\left(\bs{z}_{n-1}\right)+\Delta \bs{z}_n.
  \end{equation}
  where $\Delta\bs{z}_n=\left(\Delta x_1,...,\Delta x_N;\Delta
  p_1,\dots,\Delta p_N\right)^T$ is the round-off error vector when
  performing the $n^{\text{th}}$ iteration.  Similarly, the reversal
  integration can be written as,
  \begin{equation}
    \bs{z}_{n-1}^{\prime}=f^{-1}\left(\bs{z}_n^{\prime}\right)+\Delta
    \bs{z}_n^{\prime},
  \end{equation}
  where, $f^{-1}$ is the inverse map, and primes ($^{\prime}$) denote
  the coordinates of the reversal so as to distinguish from the
  forward trajectory.  The errors
  $\Delta\bs{z},\;\Delta\bs{z}^{\prime}$ are distributed uniformly and
  randomly within a range determined by the values of
  $\bs{z},\;\bs{z}^{\prime}$, and the number of bit of the computation
  unit~\cite{ieee754-2019}.

  When considering a case in which only one F-R iteration is computed,
  $\bs{z}_0\overset{f}{\rightarrow}\bs{z}_1
  \overset{f^{-1}}{\rightarrow}\bs{z}_0^{\prime}$. The difference
  between $\bs{z}_0$ and $\bs{z}_0^{\prime}$ can be estimated with
  local linear derivatives,
  \begin{align}\label{eq:fbDiff}
    \bs{z}_0^{\prime}-\bs{z}_0 & = f^{-1}\left(f(\bs{z}_0)+\Delta
    \bs{z}_1\right)+\Delta \bs{z}_1^{\prime}-\bs{z}_0\nonumber \\ &
    \approx \frac{\partial
      f^{-1}}{\partial\bs{z}}\biggr\rvert_{f(\bs{z}_0)}\Delta
    \bs{z}_1+\Delta \bs{z}_1^{\prime}\nonumber\\ &
    =\left[\frac{\partial
      f}{\partial\bs{z}}\biggr\rvert_{\bs{z}_0}\right]^{-1}\Delta
    \bs{z}_1+\Delta \bs{z}_1^{\prime},
  \end{align}
  where the inverse Jacobian matrix of $f$ is evaluated at
  $\bs{z}_0$. For the sake of simplicity, the linearized matrix for
  1-dimension $x$-$p$ is shown as,
  \begin{equation}\label{eq:localMat}
    \left[\frac{\partial f}{\partial\bs{z}}\biggr\rvert_{\bs{z}_0}\right]^{-1} =
      \left[\begin{array}{cc}
        \frac{\partial x_1}{\partial x_0} & 
        \frac{\partial x_1}{\partial p_0} \\ 
        \frac{\partial p_1}{\partial x_0} & 
        \frac{\partial p_1}{\partial p_0}
    \end{array}\right]^{-1}.
  \end{equation}
  Equation~\ref{eq:fbDiff} indicates that the difference
  $|\bs{z}_0^{\prime}-\bs{z}_0|$ originates from random round-off
  errors, which are scaled by an inverse Jacobian matrix
  Eq.~\ref{eq:localMat} on the passage of $\bs{z}_0$.

  The difference, $\left|\bs{z}_0^{\prime}-\bs{z}_0\right|$, from one
  iteration can sometimes be impacted by random round-off noise,
  rather than the dynamical systems themselves as desired. On the
  other hand, if chaos is sufficiently weak, the difference is still
  invisible by just one-time scaling. Therefore, to overcome this
  difficulty, it may be necessary to implement multiple iterations,
  $\bs{z}_0\overset{f^N}{\rightarrow}
  \bs{z}_N\overset{f^{-N}}{\rightarrow}\bs{z}_0^{\prime}$,
  ($N\ge2$). The difference can be estimated similarly as
  Eq.~\ref{eq:fbDiff},
  \begin{align}\label{eq:niter}
    \left(\bs{z}_0^{\prime}-\bs{z}_0\right){}_{N} & \approx
    \Delta\bs{z}_1^{\prime}+\sum_{n=2}^{N}\left(\prod_{j=0}^{n-2}\left[
      \frac{df}{d\bs{z}}\biggr\rvert_{\bs{z}=f^{j}\left(\bs{z}_0\right)}
      \right]^{-1}\right)\Delta\bs{z}^{\prime}_{n}\nonumber \\
    & + \sum_{n=1}^{N}\left(\prod_{j=0}^{n-1}\left[
      \frac{df}{d\bs{z}}\biggr\rvert_{\bs{z}=f^{j}\left(\bs{z}_0\right)}
      \right]^{-1}\right)\Delta\bs{z}_{n}.
  \end{align}
  Here $f^j$ represents the value of the $j^{\text{th}}$-iterations of
  the map $f$ without round-off error. Equation~\ref{eq:niter}
  illustrates that round-off errors are accumulated during each
  iteration and are then scaled by local linear matrices along the
  trajectories in both directions.  With sufficient iterations, the
  cumulative difference indicates the chaos of the trajectory.

  It is worth noting that, even if a system has no chaos, the
  cumulative random error between forward integration and its
  corresponding reversal is directly proportional to the number of
  iterations executed~\cite{Laslett:1957}. If chaos is present,
  however, the error will grow exponentially.
 
  In large scale modern accelerators, F-R integrations need to be
  evaluated magnet-by-magnet. A full cycle around an accelerator
  equates to one iteration as described above. The round-off errors
  $\Delta\bs{z}$ receive a contribution from each integration step. A
  short-term tracking simulation could generate an observable
  difference when 64 bit floats were used. To be specific, only
  one-turn F-R integrations are sufficient to optimize the DA of the
  NSLS-II storage ring.  Usually these differences are observable but
  still at quite a small scale. Therefore, a base-ten logarithm is used
  to allow a large range to better represent them,
  \begin{equation}\label{eq:chaos}
    \Delta = \log_{10} |\bs{z}_0-\bs{z}_0^{\prime}|.
  \end{equation}

\section{\label{sect:henon}H\'{e}non map}

  In this section, the F-R integration method is used to
  study a 1-dimensional H\'enon map,
    \begin{equation}
    \left(\begin{array}{c}
      x\\
      p
    \end{array}\right)_{n}=\left(\begin{array}{cc}
      \cos\mu & \sin\mu\\
      -\sin\mu & \cos\mu
    \end{array}\right)\left(\begin{array}{c}
      x\\
      p-x^{2}
    \end{array}\right)_{n-1}.
    \end{equation}
  This discrete H\'enon map represents a thin-lens sextupole kick
  followed by a linear phase space rotation at a phase advance
  $\mu$. Its reversal map can be expressed as an inverse rotation
  followed by an inverse thin-lens kick,
  \begin{align}
    \left(\begin{array}{c}
      x_{t}\\
      p_{t}
    \end{array}\right) & =\left(\begin{array}{cc}
      \cos\mu & -\sin\mu\\
      \sin\mu & \cos\mu
    \end{array}\right)\left(\begin{array}{c}
      x\\
      p
    \end{array}\right)_n,\nonumber \\ 
    \left(\begin{array}{c}
      x\\
      p
    \end{array}\right)_{n-1} & =\left(\begin{array}{c}
      x_{t}\\
      p_{t}+x_{t}^{2}
    \end{array}\right),
  \end{align}
  where $x_t,\;p_t$ are the intermediate variables.

\begin{figure}[!ht]
     \includegraphics[width=0.48\columnwidth]{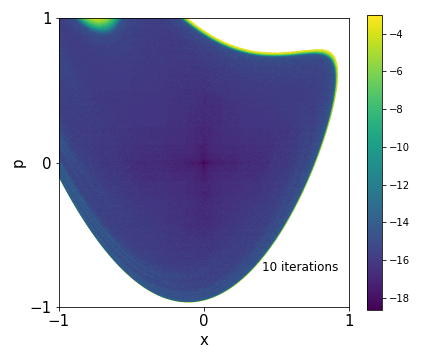}
     \includegraphics[width=0.48\columnwidth]{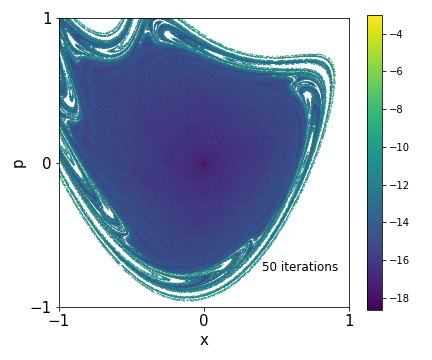}
     \includegraphics[width=0.48\columnwidth]{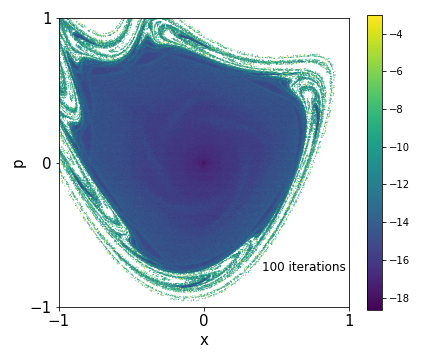}
     \includegraphics[width=0.48\columnwidth]{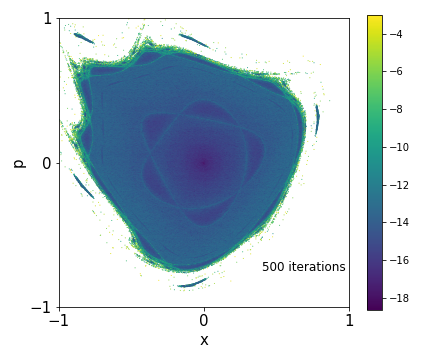}
     \caption{\label{fig:henon}(Colored) Contour of the F-R
       integrations with different numbers of iterations for a H\'enon
       map. The colormap are the difference of initial conditions as a
       function of the phase space coordinates $x$-$p$.  The white
       area represents unbounded trajectories by manually setting a
       threshold of $|x|>10$.  More iterations provide more detailed
       chaos information, but even with just a few dozen iterations,
       an early indicator of chaos can be determined.}
  \end{figure}

  The H\'enon map's linear phase advance is chosen as
  $\mu=0.205\times2\pi$ in order to observe the $5^{\text{th}}$-order
  resonance line at certain amplitudes. The difference between initial
  conditions obtained from the F-R integration is illustrated in
  Fig.~\ref{fig:henon}. When the F-R integrations are calculated with
  only 10-50 iterations (as shown in the top row), the area of the
  stable region is overestimated and the inside resonances are almost
  invisible. After 100 iterations, the resonance lines and stable
  islands become gradually visible.  More iterations can provide much
  more detailed chaos information as illustrated in the two bottom
  subplots.

\section{\label{sect:application}Applications}

  In this section we demonstrate this method by optimizing the dynamic
  apertures for the National Synchrotron Light Source II
  (NSLS-II)~\cite{NSLS-II:2013} main storage ring and a test
  diffraction-limited light source ring.

  \subsection{\label{sect:nsls-ii}NSLS-II storage ring}

\begin{figure}[!ht]
      \centering \includegraphics[width=1.\columnwidth]{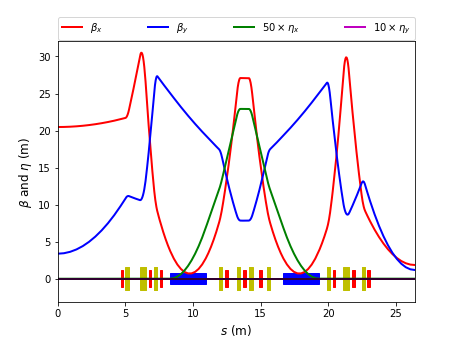}
      \caption{\label{fig:nsls2cell} (Colored) The linear optics and
        magnet layout for one of the 30 cells of the NSLS-II storage
        ring. The red blocks represent sextupoles. The three located
        between two dipoles are used to correct the natural
        chromaticity. The remaining six are used here for DA
        optimization.}
    \end{figure}

    NSLS-II is a dedicated $3^{\text{rd}}$ generation medium energy (3
    GeV) light source operated by Brookhaven National Laboratory. Its
    main storage ring's lattice is a typical double-bend-achromat
    structure. Its linear optics for one cell is illustrated in
    Fig.~\ref{fig:nsls2cell}. The whole ring is composed of 30 such
    cells. The natural chromaticities are corrected to $+2/+2$ at the
    transverse plane by the chromatic sextupoles. The optimization
    knobs are six families of harmonic sextupoles located at
    dispersion-free sections. The goal of optimization is to obtain a
    sufficient DA ($|x|>15\;\textrm{mm}, |y|>5\;\textrm{mm}$) for the
    off-axis injection at the long straight section center where
    $\beta_x= 20.5\;\textrm{m},\;\beta_y=3.4\;\textrm{m}$, and a
    $|\delta|>2.5\%$ momentum acceptance to ensure a 3 hour lifetime
    at a 500 mA beam current.

  \subsection{\label{sect:zone}Optimization objectives and results}

\begin{figure}[!ht]
      \centering \includegraphics[width=1.\columnwidth]{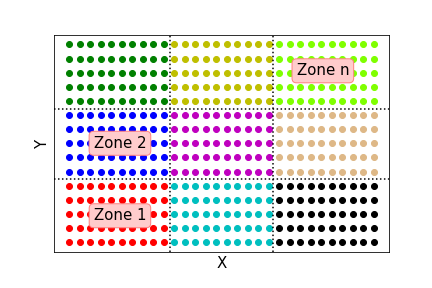}
      \caption{\label{fig:zone} (Colored) Dividing the region of
        interest (ROI) into $n$ zones in the $x$-$y$ plane. In each
        zone, multiple initial conditions (represented with
        same-colored dots) are uniformly populated. The optimization
        objectives are the difference between the initial conditions
        of F-R integrations averaged over each zone.}
    \end{figure}

    On the transverse $x$-$y$ plane at the injection point, multiple
    initial conditions are uniformly populated within a Region Of
    Interest (ROI). The ROI is chosen to cover the needed aperture.
    The virtual particle trajectories are simulated with a $4^{th}$
    order kick-drift symplectic integrator~\cite{Yoshida:1990} in
    which negative physical length elements are allowed.  The
    symplectic integration is implemented with a python code, which
    has been independently benchmarked with another reliable tracking
    simulation code \textsc{impactz}~\cite{QIANG2000434}.  After
    evolving some revolutionary periods (usually an integer number of
    turns), their reversal trajectories are computed by switching the
    sign of the coordinate $s$ and letting particles run back to
    $s=0$.  Newly re-established initial conditions deviate from the
    original ones.  A forward integration and its reversal make up a
    pair of trajectories for comparison.  A larger difference between
    a pair of initial conditions indicate a stronger chaos.  The goal
    of optimization then becomes minimizing the difference for all
    pairs of initial conditions within the ROI.  It is not practical
    or necessary to minimize so many pairs of initial conditions
    simultaneously, therefore, the ROI is divided into several zones
    as shown in Fig.~\ref{fig:zone}.  For each zone, the difference of
    initial conditions are averaged over all F-R integrations pairs.
    Then the averaged values for all zones are used as the
    optimization objectives, which need to be minimized simultaneously
    to suppress the chaos inside the whole ROI. The optimization
    objective functions $g$ reads as
    \begin{equation}\label{eq:obj}
      \bar{\Delta}_i = g_i(K_{2,j}),
    \end{equation}
    where, $i,\;j$ are the indices of the ROI zones and the sextupoles
    respectively, $\bar{\Delta}_i$ is the average difference in the
    $i^{th}$ zone, and $K_{2,j}$ is the $j^{th}$ sextupole's
    normalized gradient.

    Quantitatively, the difference in Eq.~\ref{eq:chaos} and
    ~\ref{eq:obj} for a pair of initial conditions in the normalized
    phase space can be computed as
    \begin{equation}\label{eq:delta}
      \Delta=\log_{10}\sqrt{\Delta\bar{x}^2+\Delta\bar{p}_x^2+
        \Delta\bar{y}^2+\Delta\bar{p}_y^2},
    \end{equation}
    where $\bar{x},\;\bar{p}_x;\;\bar{y},\;\bar{p}_y$ are the
    difference of canonical coordinates normalized with Courant-Snyder
    parameters as follows~\cite{Courant:1958},
    \begin{align}\label{eq:csn}
      \begin{bmatrix}
        \Delta\bar{u}\\ 
        \Delta\bar{p}_u
      \end{bmatrix}
      =
      \begin{bmatrix}
        \frac{1}{\sqrt{\beta_u}} & 0\\
        \frac{\alpha_u}{\sqrt{\beta_u}} & \sqrt{\beta_u}
       \end{bmatrix}
      \begin{bmatrix}
        \Delta u\\
        \Delta p_u
      \end{bmatrix}.
    \end{align}
    where $u=x,\;\text{or}\;y$, The normalization of
    Eq.~\ref{eq:csn} expresses the canonical coordinate pairs in the
    same units $m^{1/2}$ for arithmetic addition.

\begin{figure}[!ht]
      \centering \includegraphics[width=1.\columnwidth]{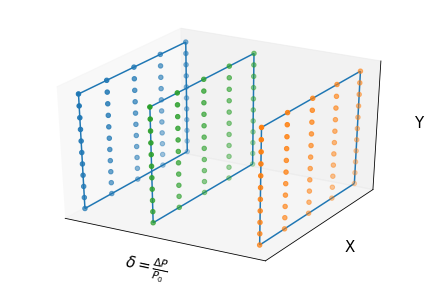}
      \caption{\label{fig:zone_dp} (Colored) Optimizing several fixed
        off-momentum DA simultaneously. By separating a 5-dimensional
        phase space ($x,p_x;y,p_y;\delta$) into several slices along
        the $\delta$-axis, DA for off-momentum particles can be
        optimized simultaneously.}
    \end{figure}

    To obtain sufficient beam lifetime and DA simultaneously, one must
    optimize them simultaneously~\cite{Borland:2015}. Direct
    optimization of beam lifetime is time-consuming. An alternative is
    to optimize different off-momentum DA. This was achieved by a
    $\delta$-slicing method as illustrated in
    Fig.~\ref{fig:zone_dp}. First, the desired energy acceptance range
    is determined based on the beam scattering lifetime calculation at
    a certain beam current. Then several sliced off-momentum DA are
    included into the optimization objectives. At each slice, the
    objective functions are evaluated in the same way as
    Fig.~\ref{fig:zone}.

\begin{figure}[!ht]
      \centering \includegraphics[width=1.\columnwidth]{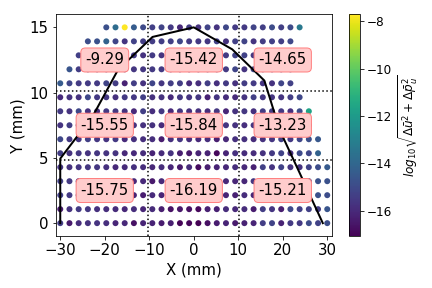}
      \caption{\label{fig:zone_obj} (Colored) The objective function
        evaluated in 9 zones for the $\delta=0$ slice.  These were
        obtained for a specific set of sextupoles settings for the
        NSLS-II storage ring. Blank points represent lost particles
        ($|x,y|>1$ m) within 1 turn tracking. The maximum allowed
        number of lost particles is used as an optimization
        constraint. The black line is the dynamic aperture obtained by
        multi-turn (1,024) tracking simulation with the code
        \textsc{elegant}. The one-turn F-R integrations give a more
        optimistic result than the multi-turn tracking simulation.  As
        an early indicator of chaos, however, it does provide a
        reasonable criteria for the optimizer.}
    \end{figure}

    Multiple zones within the ROI for different momentum slices need
    to be minimized simultaneously. The multi-objective genetic
    algorithm (MOGA) was used for this task.  More turns of particle
    tracking simulation can indicate the chaos more accurately, but
    this requires more computation time. After manually checking the
    dependence of the chaos indicator against the number of turns,
    one-turn F-R integration (crossing 30 cells) was chosen to compute
    this early chaos indicator as illustrated in
    Fig.~\ref{fig:zone_obj}.  Although the early indicator of chaos
    from F-R integration provides an optimistic approximation, it does
    rule out many of the less competitive candidates and narrows down
    the parameter search range quickly.  By allowing a small-scale
    population, which includes the evolution of only 1,000 candidates
    over just 50 generations, the top candidates' average fitness is
    seen to converge. It took about 6 hours to complete the
    optimization with 50 Intel\textsuperscript{\textregistered}
    Xeon\textsuperscript{\textregistered} 2.2-2.3 GHz CPU cores.
    Another reliable tracking code
    \textsc{elegant}~\cite{Borland:2000gvh} was then used to check the
    DA for all the candidates only inside the last generation. Among
    them, some of the elite candidates are selected for more extensive
    simulation studies to check their final performance.
    
\begin{figure}[!ht]
      \centering \includegraphics[width=1.\columnwidth]{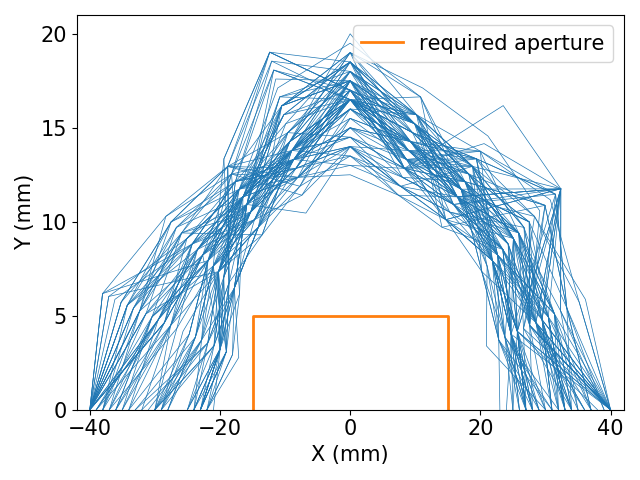}
      \caption{\label{fig:moga} (Colored) DA of the top 100 candidates
        (measured with the area) from the $50^{th}$ generation of the
        evolved population obtained with the MOGA optimizer. The light
        yellow box is the required aperture for the off-axis top-off
        injection.}
    \end{figure}

    The DA profiles for the top 100 candidates inside the last
    generation are illustrated in Fig.~\ref{fig:moga}. Although the
    six sextupole families settings are very different, their DA
    satisfy the minimum requirement for top-off injection.  This
    observation confirms that short-term F-R integration can indeed be
    used for DA optimization.  Among these candidates, one from the
    elite cluster was selected to carry out a more detailed frequency
    map analysis (FMA) to verify its nonlinear dynamics
    performance. The FMA results are summarized later in
    Sect.~\ref{subsect:fma}.
    
    In the specific example of a dedicated light source machine (the
    NSLS-II storage ring), the longitudinal synchrotron oscillation
    has not been included.  It is straightforward to include it if
    needed. It can be done by extending Eq.~\ref{eq:delta} to the
    6-dimensional space when the betatron-synchrotron coupling
    resonances become critically important, e.g. in the case of
    collider rings.

  \subsection{Comparison with frequency map analysis}\label{subsect:fma}

    The frequency map analysis (FMA) is widely used to evaluate the
    performance of a nonlinear lattice~\cite{Robin:2000, Laskar:2003,
      Papaphilippou:2014, Todesco:1996}. The FMA was also applied
    directly to optimizing the DA of a light source
    ring~\cite{Steier:2010}. By comparing the tune diffusion rate
    determined by two pieces of turn-by-turn simulation or measurement
    data, the resonances of the lattice can be visualized. In our
    example, one elite solution was selected from the last generation
    of candidates to carry out a detailed FMA to characterize its
    nonlinear dynamics performance. In the meantime, a multi-turn
    (1,024 turns) F-R analysis was conducted for a comparison with the
    FMA results. The sextupole settings for the current NSLS-II
    lattice and the selected elite solution are listed in
    Tab.~\ref{tab:sextK2} for comparison.

    \begin{table}[!ht]
      \centering
      \caption{\label{tab:sextK2}Comparison of two sextupoles settings}
      \begin{tabular}{p{2.2cm} | p{1.2cm} p{2.2cm} p{2.2cm}} 
        \hline
        Sextupole & unit & $K_2$ (current) & $K_2$ (F-R)  \\
        \hline\hline
        SH1 & $\textrm{m}^{-3}$ & 19.8329 &  19.8495 \\ 
        SH3 & $\textrm{m}^{-3}$ & -5.8551 &  -0.4017 \\ 
        SH4 & $\textrm{m}^{-3}$ &-15.8209 & -22.0160 \\ 
        SL3 & $\textrm{m}^{-3}$ &-29.4609 & -29.0057 \\ 
        SL2 & $\textrm{m}^{-3}$ & 35.6779 &  27.9185 \\ 
        SL1 & $\textrm{m}^{-3}$ &-13.2716 &  -2.6051 \\ 
        \hline
      \end{tabular}
    \end{table}

\begin{figure}[!ht]
      \centering \includegraphics[width=1.\columnwidth]{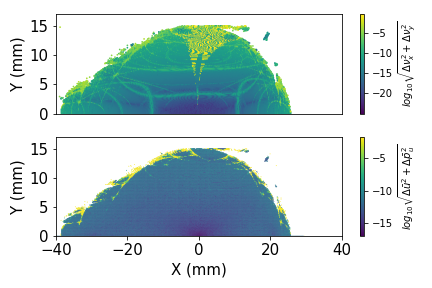}
      \caption{\label{fig:fmafb} (Colored) Top: FMA on the $x$-$y$
        plane for 1,024 turns of data (512 leading and 512 trailing
        turns) with the code \textsc{elegant}. Bottom: F-R analysis
        for 1,024 turns. Using the FMA, some unusual diffusion rate
        (as shown in yellow stripes near $x=0$) can be observed.} 
    \end{figure}

\begin{figure}[!ht]
      \centering \includegraphics[width=1.\columnwidth]{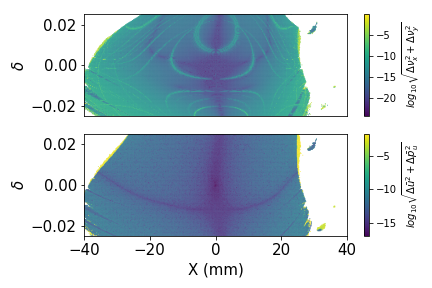}
      \caption{\label{fig:fmafb_dp} (Colored) Top: FMA on the
        $x$-$\delta$ plane for 1,024-turns of data (512 leading and
        512 trailing turns) with the code \textsc{elegant}. Bottom:
        F-R analysis for 1,024 turns.}
    \end{figure}

    Figure~\ref{fig:fmafb} illustrates the on-momentum DA in the
    transverse $x$-$y$ plane. Figure~\ref{fig:fmafb_dp} shows the
    off-momentum acceptance in the $x$-$\delta$ plane. The FMA results
    yield more copious and fine patterns of the resonance than the FRI
    results.  Even the accuracy of the NAFF can be theoretically
    proportional to $1/N^4$ ($N$ is the total number of the sampling data)
    with a Hanning window~\cite{Laskar:2002bk}, it is still relatively
    slower compared to the exponential improvement of the reversal
    method. Therefore, a much less number of the FRI tracking is needed to
    drive the optimizer. In the previous example, only one-turn FRI has
    been used to speed up the convergence.

\begin{figure}[!ht]
      \centering \includegraphics[width=1.\columnwidth]{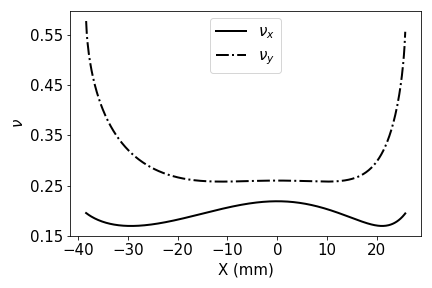}
      \caption{\label{fig:tswa} Tune shift with initial coordinate in
        the horizontal plane for the selected candidate. Although the
        vertical tune rises suddenly faster at larger horizontal
        amplitudes, the tune variations are within $\pm0.03$ inside
        the range of $x\in[-15,15]\;\textrm{mm}$.}
    \end{figure}

\begin{figure}[!ht]
      \centering \includegraphics[width=1.\columnwidth]{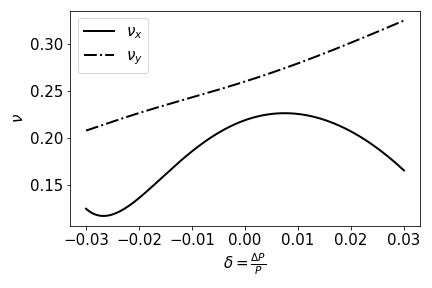}
      \caption{\label{fig:chrom} Tune variation with the momentum
        offset, i.e. chromaticity for the selected candidate. The
        linear chromaticities were tuned to +2 for both $x$ and $y$
        planes.}
    \end{figure}

\begin{figure}[!ht]
      \centering \includegraphics[width=1.\columnwidth]{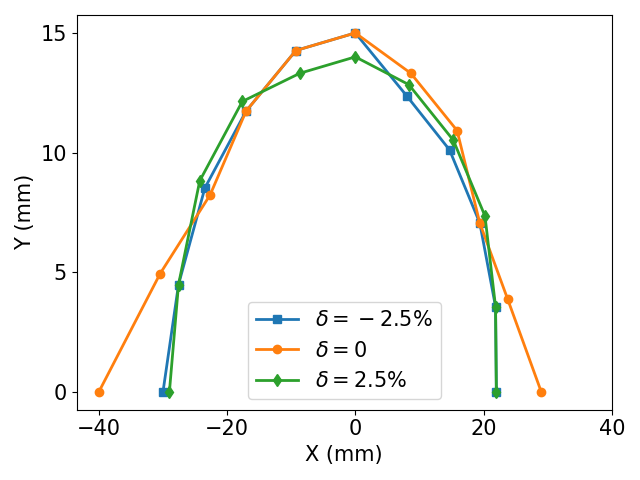}
      \caption{\label{fig:on_off_DA} (colored) On- and two off-momentum
        ($\delta=\pm2.5\%$) DAs for the selected candidate.}
    \end{figure}

    A control of higher order chromaticities and
    amplitude-dependent-tune-shifts to avoid destructive
    resonance-crossing is critical in DA optimization. This could be
    achieved by minimizing some specific nonlinear driving
    terms~\cite{Dragt:2011vea, Chao:2002st}. For example, $C_{2200,0},
    C_{0022,0}, C_{1111,0}$ are the first order
    amplitude-dependent-tune-shift coefficients; then $C_{1100,n},
    C_{0011,n}, n\ge2$ are the higher order chromaticity
    coefficients. These terms can be used as either objective
    functions or explicit constraints.  In the F-R integration method,
    no explicit constraints are used to limit them.  The final
    tracking simulation on the selected solution, however, shows that
    both the amplitude-dependent-tune-shifts (Fig.~\ref{fig:tswa}) and
    the higher order chromaticities (Fig.~\ref{fig:chrom}) are
    automatically and passively suppressed. The on-momentum and two
    off-momentum ($\pm2.5\%$) DA, computed with the code
    \textsc{elegant}, are shown in Fig.~\ref{fig:on_off_DA}.

    The optimization was implemented on an error-free model. Then the
    systematic and random magnetic field errors and the misalignments
    have been included to confirm the robustness of the solution. An
    online beam test on the NSLS-II storage ring was also carried out
    to confirm that the off-axis top-off injection efficiency is
    between 95-100\%, which is comparable with our current operation
    lattice. The beam lifetime at a 400 mA beam current is longer than
    5.5 hour, with a diffraction-limited vertical beam emittance of 8
    pm. While the current operation lattice was observed having 4.5
    hours in similar conditions.

  \subsection{MBA lattice for diffraction-limited light source}

\begin{figure}[!ht]
      \centering \includegraphics[width=1.\columnwidth]{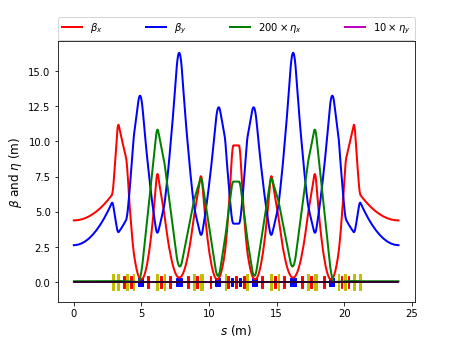}
      \caption{\label{fig:hals_optics} (Colored) Linear optics and
        magnet layout for one cell of a test MBA lattice.}
    \end{figure}

\begin{figure}[!ht]
      \centering \includegraphics[width=1.\columnwidth]{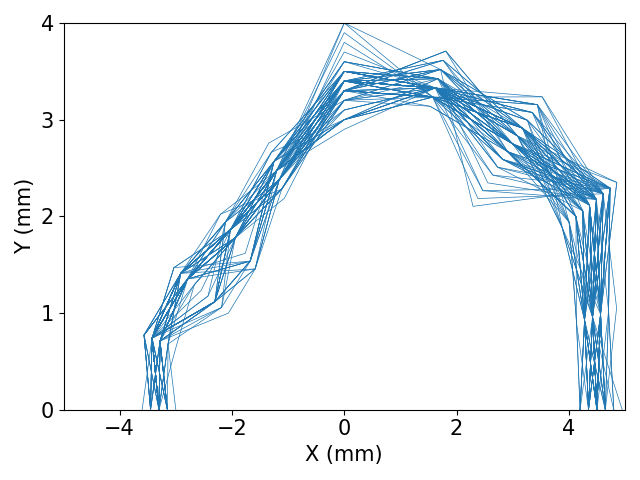}
      \caption{\label{fig:hals} (Colored) On-momentum DA for the top
        100 candidates for the test MBA lattice.}
    \end{figure}

    The F-R integration method has also been used to test on a
    multi-bend-achromat (MBA) structure, which could potentially be
    used as a diffraction-limited light source storage ring lattice in
    the future. The horizontal emittance of the test MBA lattice used
    was 78 pm at a beam energy of 2 GeV. The linear lattice is shown
    in Fig.~\ref{fig:hals_optics}, in which most sextupoles are
    chromatic sextupoles.  The MOGA result showing the top 100
    candidates’ apertures are illustrated in Fig.~\ref{fig:hals}. The
    preliminary result confirms that the F-R integration could also be
    applied to a more complicated nonlinear lattice, and the approach
    itself should be general in optimizing other lattices.

\section{\label{sect:summary}Summary}

  An indicator of chaos obtained with forward-reversal integration has
  been used for optimization of dynamic aperture of storage rings. The
  indicator, intrinsically but empirically associated with the
  Lyapunov exponent, gives an early indication of the chaos of beam
  motion in storage rings. Although the indicator cannot give the
  exact dynamic aperture profile with a short-term tracking
  simulation, a concrete correlation and large MOGA candidate pool
  yield some optimal lattice solutions. The NSLS-II storage ring and a
  test MBA lattice are used as examples to illustrate the application
  of this method.

  The computation of the difference of F-R integrations has been
  implemented in the \textsc{elegant}~\cite{Borland:2000gvh} code since
  version 2019.4.0. Besides the F-R integration, the \textsc{elegant} code
  also provides another option for the users to compute the change in
  linear actions $J_{x,y}$ from two forward-only trackings, with a small
  difference in their initial conditions. By properly choosing small
  changes based on the machine precision, the signs and absolute values of
  initial conditions, and the round-off method in computer operation
  system, one should be able to get a similar result as the F-R
  integration.  However, the needed implementation in this option is more
  complicated than the F-R integration. The fundamental principle of this
  method is to numerically characterize the sensitivity of a chaotic
  motion to its initial conditions by using round-off errors.

\section*{Acknowledgements}
  We would like to thank C. Mitchell and R. Ryne (LBL) for the
  stimulating and collaborative discussions, J. Qiang (LBL) for
  providing the \textsc{impactz} code, Z. Bai (USTC) for providing the
  test MBA lattice, M. Giovannozzi (CERN) for fruitful discussions and
  constructive suggestions, M. Borland (ANL) for implementing this
  method in the \textsc{elegant} code, and I. Morozov (BINP) for
  pointing out a numerical error in the manuscript.  This research
  used resources of the National Synchrotron Light Source II, a
  U.S. Department of Energy (DOE) Office of Science User Facility
  operated for the DOE Office of Science by Brookhaven National
  Laboratory (BNL) under Contract No. DE-SC0012704, and the computer
  resources at the National Energy Research Scientific Computing
  Center. This work was also supported by (1) the Accelerator
  Stewardship program under the Office of High Energy Physics; (2)
  Lawrence Berkeley National Laboratory operated for the DOE Office of
  Science under Contract No. DE-AC02-05CH11231, (3) BNL's Laboratory
  Directed Research and Development program ``NSLS-II High Brightness
  Upgrade and Design Studies'' No. 17-015. (4) DOE SBIR grant under
  Contract No. DE-SC0019538. One author (KH) acknowledges the support
  from the U.S. DOE Early Career Research Program under the Office of
  High Energy Physics.
  
\bibliography{fb_dyap.bib}

\end{document}